\documentclass{article}
\usepackage{spconf2027,amsmath,amssymb,graphicx,booktabs,microtype,placeins,hyperref}
\usepackage{fix-cm}
\usepackage{tikz}
\usetikzlibrary{arrows.meta}
\hypersetup{hidelinks}
\title{Alignment-Path Distillation from Non-streaming ASR-LLMs for Streaming Speech Recognition}
\name{Yan Jia, Kai Huang, Junjie Chen, Feng-Long Xie, Xu Tang, Yao Hu}
\address{Xiaohongshu Inc.}
\begin{document}
\raggedbottom
\setlength{\abovedisplayskip}{4pt plus 2pt minus 1pt}
\setlength{\belowdisplayskip}{4pt plus 2pt minus 1pt}
\setlength{\abovedisplayshortskip}{3pt plus 1pt minus 1pt}
\setlength{\belowdisplayshortskip}{3pt plus 1pt minus 1pt}
\setlength{\textfloatsep}{10pt plus 2pt minus 2pt}
\setlength{\dbltextfloatsep}{10pt plus 2pt minus 2pt}
\setlength{\intextsep}{10pt plus 2pt minus 2pt}
\maketitle
\begin{abstract}
In this paper, we propose an alignment-path distillation framework for streaming automatic speech recognition (ASR) with large language models (LLMs). Interleaved streaming ASR-LLMs use forced alignments (FA) from alignment models, such as those trained with connectionist temporal classification (CTC), to construct speech--text training sequences. However, alignments obtained from a separate acoustic model may be inconsistent with those learned by LLM-based ASR. This motivates us to transfer alignment information from a non-streaming ASR-LLM to improve streaming recognition. Specifically, we extract monotonic alignment paths from a non-streaming teacher's soft text--audio attention and use them to construct interleaved training sequences. The framework also includes logit and hidden-state distillation to learn from the teacher's output distributions and internal representations. Experimental results show that, without logit or hidden-state distillation, training with teacher-derived alignment paths achieves a 5.2\% relative error rate reduction compared with training using forced alignments. When both models use logit and hidden-state distillation, teacher-derived alignments yield a 3.9\% relative error rate reduction, with similar mean emission latency but higher flicker. The complete framework achieves a 16.6\% relative error rate reduction compared with training using forced alignments without logit or hidden-state distillation.
\end{abstract}

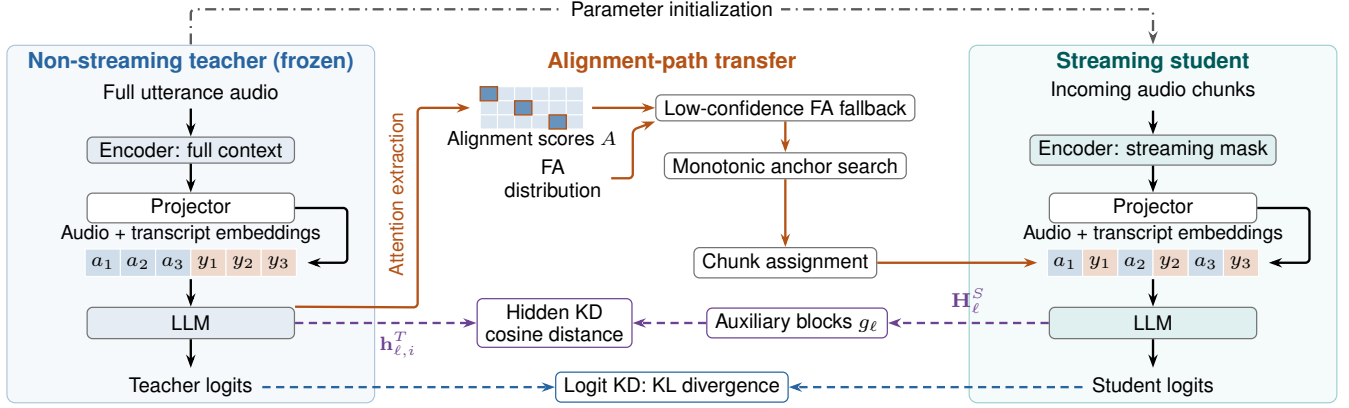
\begin{figure*}[!t]
\centering
\resizebox{\textwidth}{!}{%
\begin{tikzpicture}[x=1cm,y=1cm,font=\sffamily\fontsize{7.3}{8.3}\selectfont,
 box/.style={draw=black!55,line width=.45pt,rounded corners=2pt,fill=white,align=center,inner xsep=3pt,inner ysep=2pt},
 arr/.style={-{Stealth[length=1.7mm,width=1.2mm]},line width=.85pt,rounded corners=1.5pt},
 kd/.style={arr,dash pattern=on 3pt off 2pt},
 lab/.style={font=\sffamily\bfseries\fontsize{8}{9}\selectfont}]
\definecolor{pathcol}{RGB}{180,82,22}
\definecolor{predcol}{RGB}{39,100,160}
\definecolor{hidcol}{RGB}{109,67,145}
\path[use as bounding box] (0,.35) rectangle (17.2,5.65);
\draw[arr,black!65,dash pattern=on 4pt off 2pt on .7pt off 2pt]
 (2.45,5.03)--(2.45,5.48)--(14.75,5.48)--(14.75,5.03);
\node[fill=white,inner sep=3pt,align=center,font=\sffamily\fontsize{7.3}{8.3}\selectfont]
 at (8.6,5.48) {Parameter initialization};
\draw[rounded corners=3pt,draw=predcol!40,fill=predcol!4] (.1,.45) rectangle (4.8,5.02);
\draw[rounded corners=3pt,draw=teal!40,fill=teal!4] (12.4,.45) rectangle (17.1,5.02);
\node[lab,text=predcol] at (2.45,4.80) {Non-streaming teacher (frozen)};
\node[lab,text=teal!70!black] at (14.75,4.80) {Streaming student};
\node (at) at (2.45,4.43) {Full utterance audio};
\node (as) at (14.75,4.43) {Incoming audio chunks};
\node[box,minimum width=2.65cm,fill=predcol!12] (ct) at (2.45,3.68) {Encoder: full context};
\node[box,minimum width=2.65cm,fill=teal!12] (cs) at (14.75,3.68) {Encoder: streaming mask};
\node[box,minimum width=2.65cm] (pt) at (2.45,2.94) {Projector};
\node[box,minimum width=2.65cm] (ps) at (14.75,2.94) {Projector};
\draw[arr] (at.south)--(ct.north);\draw[arr] (ct.south)--(pt.north);
\draw[arr] (as.south)--(cs.north);\draw[arr] (cs.south)--(ps.north);
\node[font=\sffamily\fontsize{7}{8}\selectfont] at (2.45,2.60) {Audio + transcript embeddings};
\node[font=\sffamily\fontsize{7}{8}\selectfont] at (14.75,2.60) {Audio + transcript embeddings};
\foreach \x/\t/\c in {1.1/a_1/predcol,1.55/a_2/predcol,2/a_3/predcol,2.45/y_1/pathcol,2.9/y_2/pathcol,3.35/y_3/pathcol}{
\node[fill=\c!22,draw=white,minimum width=.45cm,minimum height=.35cm,inner sep=0pt] at (\x+.225,2.24) {$\t$};}
\foreach \x/\t/\c in {13.4/a_1/predcol,13.85/y_1/pathcol,14.3/a_2/predcol,14.75/y_2/pathcol,15.2/a_3/predcol,15.65/y_3/pathcol}{
\node[fill=\c!22,draw=white,minimum width=.45cm,minimum height=.35cm,inner sep=0pt] at (\x+.225,2.24) {$\t$};}
\draw[arr] (pt.east)--(4.45,2.94)--(4.45,2.24)--(3.94,2.24);
\draw[arr] (ps.east)--(16.75,2.94)--(16.75,2.24)--(16.23,2.24);
\node[box,minimum width=2.65cm,minimum height=.40cm,fill=predcol!12] (qt) at (2.45,1.47) {LLM};
\node[box,minimum width=2.65cm,minimum height=.40cm,fill=teal!12] (qs) at (14.75,1.47) {LLM};
\draw[arr] (2.45,2.03)--(qt.north);
\draw[arr] (14.75,2.03)--(qs.north);
\node (lt) at (2.45,.65) {Teacher logits};
\node (ls) at (14.75,.65) {Student logits};
\draw[arr] (qt.south)--(lt.north);
\draw[arr] (qs.south)--(ls.north);
\node[lab,text=pathcol] at (8.6,4.80) {Alignment-path transfer};
\foreach \i in {0,...,2}{\foreach \j in {0,...,5}{
\filldraw[fill=predcol!17,draw=white,line width=.25pt] (6.15+.22*\j,4.49-.18*\i) rectangle +(.22,-.18);}}
\foreach \i/\j in {0/0,1/2,2/4}{
\filldraw[fill=predcol!70,draw=pathcol,line width=.5pt] (6.15+.22*\j,4.49-.18*\i) rectangle +(.22,-.18);}
\node[font=\sffamily\fontsize{7}{8}\selectfont] at (6.81,3.81) {Alignment scores $A$};
\node[box] (fa) at (10.05,4.22) {Low-confidence FA fallback};
\draw[arr,pathcol] (7.58,4.22)--(fa.west);
\draw[arr,pathcol] (qt.north east)--(5.35,1.67)--(5.35,4.22)--(6.02,4.22);
\node[rotate=90,text=pathcol,inner sep=1pt,font=\sffamily\fontsize{7}{8}\selectfont] at (5.08,3.12) {Attention extraction};
\node[box] (mas) at (10.05,3.48) {Monotonic anchor search};
\draw[arr,pathcol] (fa.south)--(mas.north);
\node[align=center] (fain) at (7.08,3.32) {FA\\distribution};
\draw[arr,pathcol] (fain.east)--(8.1,3.32)--(8.1,3.94)--(fa.south west);
\node[box] (assign) at (10.05,2.24) {Chunk assignment};
\draw[arr,pathcol] (mas.south)--(assign.north);
\draw[arr,pathcol] (assign.east)--(13.27,2.24);
\node[box,draw=hidcol] (hk) at (7.1,1.47) {Hidden KD\\cosine distance};
\node[box,draw=hidcol] (aux) at (10.2,1.47) {Auxiliary blocks $g_\ell$};
\draw[kd,hidcol] (qs.west)--node[above] {$\mathbf H^S_\ell$}(aux.east);
\draw[kd,hidcol] (aux.west)--(hk.east);
\draw[kd,hidcol] (qt.east)--(hk.west);
\node[text=hidcol,font=\sffamily\fontsize{7}{8}\selectfont] at (5.10,1.19) {$\mathbf h^T_{\ell,i}$};
\node[box,draw=predcol] (lk) at (8.6,.65) {Logit KD: KL divergence};
\draw[kd,predcol] (lt.east)--(lk.west);
\draw[kd,predcol] (ls.west)--(lk.east);
\end{tikzpicture}}
\caption{Framework overview. The dash-dotted arrow denotes student initialization. FA fallback and monotonic search determine audio--text interleaving. Dashed arrows denote logit/hidden KD. The teacher and auxiliary blocks are used only during training. Alignment scores are illustrative.}
\label{fig:overview}
\end{figure*}

\section{Introduction}
Recent studies have explored streaming recognition with LLM-based ASR systems and related decoder-only architectures. Speech ReaLLM interleaves speech and text embeddings using alignments from an external CTC model and predicts BLANK to wait for the next speech embedding~\cite{seide2024}. SpeechLLM-XL processes audio in chunks and segments training transcripts using FA from the encoder's CTC outputs~\cite{jia2024}. BESTOW uses a streaming read/write policy~\cite{chen2024}. In a related decoder-only ASR architecture, Tsunoo et al. combine CTC compression with prefix training~\cite{tsunoo2024}. Uni-ASR trains one LLM-based model for both streaming and non-streaming recognition~\cite{xia2026}. In interleaved systems, the assignment of text tokens to audio chunks determines the acoustic context available for each token prediction.

Limited audio context remains a challenge for streaming recognition. Knowledge distillation (KD) transfers supervision from a non-streaming teacher through generated transcripts~\cite{doutre2020}, soft output distributions~\cite{hinton2015,kurata2020}, and hidden representations~\cite{kojima2021,shim2023}. Joint streaming and non-streaming training also transfers full-context knowledge to streaming recognition~\cite{yu2020}. Beyond these targets, several studies investigate temporal alignment in distillation by aligning transducer posterior peaks~\cite{kurata2020}, accounting for time shifts between streaming and non-streaming outputs~\cite{weninger2022,li2025}, transferring CTC token boundaries to monotonic attention~\cite{inaguma2021}, or distilling output distributions along teacher alignment paths~\cite{yang2022} and alignment distributions themselves~\cite{chang2023}. These studies motivate transferring alignment knowledge in addition to predictions and representations. For interleaved ASR-LLMs, we use the teacher's own text--audio attention to construct streaming training sequences, whose token-to-chunk assignments are not determined by logit or hidden-state losses alone.

However, alignments obtained from a separate acoustic model may be inconsistent with those learned by LLM-based ASR. In this paper, we propose an alignment-path distillation framework that uses alignment information from a non-streaming ASR-LLM teacher to improve streaming recognition. We extract text--audio attention from the teacher, replace low-confidence attention distributions with FA distributions, and obtain monotonic alignment paths to construct the student's interleaved training sequences. On this training sequence, logit and hidden-state distillation provide additional supervision from the teacher's output distributions and internal representations. Hidden-state distillation uses auxiliary decoder blocks to transform student representations before comparison with the teacher representations. The teacher and auxiliary blocks are used only during training. Inference uses the streaming student. Compared with training using FA, teacher-aligned training achieves relative error rate reductions of 5.2\% without logit or hidden-state distillation and 3.9\% with both losses, with similar mean emission latency but higher flicker in the latter comparison. The complete framework achieves a 16.6\% relative error rate reduction over FA-based training without either loss.

\section{Method}
Figure~\ref{fig:overview} shows the interleaved ASR-LLM and our alignment-path distillation framework.
\subsection{Interleaved ASR-LLM Architecture and Training}
The speech encoder extracts acoustic features, which the projector downsamples and maps to the LLM embedding space. The model processes consecutive audio chunks while retaining preceding audio and text as context.

Following the interleaved training paradigm of Uni-ASR~\cite{xia2026}, we divide speech into fixed-duration chunks and segment the reference transcript according to token alignment timestamps. Each audio chunk is followed by its assigned reference text. With audio embeddings $\mathbf a_k$ and text embeddings $\mathbf y^{(k)}$, the LLM input is
\begin{equation}
 [\mathbf{a}_1,\mathbf{y}^{(1)},\mathbf{a}_2,\mathbf{y}^{(2)},\ldots,\mathbf{a}_K,\mathbf{y}^{(K)}],
\end{equation}
where $K$ is the number of chunks. Text segments may be empty.

The LLM predicts each text token conditioned on the available audio embeddings and preceding reference tokens. The cross-entropy loss $\mathcal L_{\mathrm{CE}}$ is computed over the target text tokens. Chunk-restricted encoder attention and causal LLM attention prevent access to subsequent audio chunks.

During decoding, the embeddings of each incoming audio chunk are appended to the audio--text history of each beam hypothesis after encoding and projection. The LLM extends the hypotheses autoregressively using beam search, with a token limit per chunk. At each output update, we display the current highest-scoring hypothesis. Changes in the best hypothesis can revise previously displayed text, resulting in flicker. Audio processing and text generation alternate as new chunks arrive.

\subsection{Alignment-path Distillation Framework}
The non-streaming model used to initialize the student is kept fixed as the teacher. The teacher-derived alignment path is used to construct the student's interleaved training sequence, with logit and hidden-state losses providing additional supervision.

\subsubsection{Alignment-path distillation}
\noindent\textit{Attention extraction.}
We extract each text token's attention scores over audio embeddings from the non-streaming ASR-LLM teacher, using its final LLM layer and averaging across all attention heads. Applying softmax over the audio positions yields an alignment matrix $A\in\mathbb R^{U\times F}$, where $U$ and $F$ are the numbers of text and audio tokens, respectively. The text tokens correspond to the modeling units of the LLM. Each row $A_{i,:}$ gives the attention distribution of text token $i$ over audio tokens.

\noindent\textit{Confidence-based FA fallback.}
Let $A^{\mathrm{FA}}_{i,:}$ denote the FA alignment distribution over audio tokens for text token $i$. When the peak attention probability $s_i=\max_j A_{ij}$ over audio positions falls below $\theta_{\mathrm{FA}}$, we replace its attention distribution with the FA distribution:
\begin{equation}
 \widetilde A_{i,:}=\begin{cases}
 A^{\mathrm{FA}}_{i,:},&s_i<\theta_{\mathrm{FA}},\\
 A_{i,:},&\text{otherwise}.
 \end{cases}
\label{eq:fallback}
\end{equation}

\noindent\textit{Global monotonic anchor search.}
Independent peak selection can produce non-monotonic anchors. Applying a cumulative maximum afterwards may propagate an early mistake to subsequent units. Let $e_i$ denote the audio position index assigned to text unit $i$, and let $\mathbf e=(e_0,\ldots,e_{U-1})$ denote the alignment path. Inspired by monotonic alignment search in Glow-TTS~\cite{kim2020}, we optimize all anchors jointly using the updated alignment matrix $\widetilde A$:
\begin{equation}
\mathbf e^*=\arg\max_{0\le e_0<\cdots<e_{U-1}<F}
\sum_{i=0}^{U-1}\log(\widetilde A_{i,e_i}+\epsilon),
\label{eq:search}
\end{equation}
where $\epsilon=10^{-9}$. The search selects one acoustic anchor per text unit and allows acoustic positions to be skipped. It differs from a duration path that assigns every acoustic frame to a text unit.

Let $V(i,j)$ be the largest sum of log-alignment probabilities for text units $0,\ldots,i$ when unit $i$ is assigned to audio position $j$. The recurrence is
\begin{equation}
 V(i,j)=\log(\widetilde A_{i,j}+\epsilon)+\max_{k<j}V(i-1,k).
\end{equation}
Prefix maxima and their indices allow this recurrence to be evaluated in $O(UF)$ time. Backtracking from the best last-row position recovers the anchor sequence.

Audio position indices are converted to timestamps using the time step of the projected features. These timestamps determine the text segments assigned to the fixed-duration audio chunks in the interleaved training sequence.

\subsubsection{Logit and Hidden-State Distillation}

\noindent\textit{Logit supervision.}

The student and teacher use the same LLM architecture, tokenizer, and target transcript, so their target tokens correspond one-to-one despite different speech--text sequence layouts. Let $\mathcal P$ contain the corresponding prediction-position pairs for all target tokens and $N=|\mathcal P|$. Following temperature-scaled KD~\cite{hinton2015}, we minimize
\begin{equation}
\mathcal L_{\mathrm{logit}}=\frac{\tau^2}{N}\sum_{i\in\mathcal P}
 D_{\mathrm{KL}}\left(p_i^{T,\tau}\,\|\,p_i^{S,\tau}\right),
\end{equation}
where $p^{T,\tau}$ and $p^{S,\tau}$ are teacher and student softmax distributions at temperature $\tau$. 

\noindent\textit{Hidden-state supervision.}
Hidden KD constrains internal representations at corresponding target-token prediction positions. Each selected LLM layer has its own auxiliary decoder block $g_\ell$. We compute
\begin{equation}
\mathcal L_{\mathrm{hid}}=\frac{1}{|\mathcal S|N}
\sum_{\ell\in\mathcal S}\sum_{i\in\mathcal P}
\left[1-\cos\left(g_\ell(\mathbf H^S_\ell)_i,\mathbf h^T_{\ell,i}\right)\right],
\end{equation}
where $\mathcal S$ denotes the selected LLM layers. Auxiliary blocks use bidirectional self-attention over the student sequence, masking padding only. They can access later audio and text positions during training but are removed at inference, preserving the causal student path.

The complete training objective on the path-derived sequence is
\begin{equation}
\mathcal L=\mathcal L_{\mathrm{CE}}+\lambda\mathcal L_{\mathrm{logit}}+h\mathcal L_{\mathrm{hid}}.
\end{equation}
The teacher and auxiliary blocks are used only during training. Inference uses the student model.

\begin{table*}[!t]
\centering
\caption{Recognition error rates (\%): CER for Chinese and WER for English (LibriSpeech). The non-streaming teacher uses full-utterance audio and is included for reference. All is the reported aggregate error rate. Bold marks the best streaming result in each column.}
\label{tab:cer}
\renewcommand{\arraystretch}{0.92}
\begin{tabular}{lrrrrrrr}
\toprule
 & AISHELL & KeSpeech & \multicolumn{2}{c}{WeNetSpeech} & \multicolumn{2}{c}{LibriSpeech} & All\\
\cmidrule(lr){4-5}\cmidrule(lr){6-7}
System & test & test & test-net & test-meeting & test-clean & test-other & \\
\midrule
Teacher (non-streaming) & 1.09 & 3.57 & 5.05 & 4.71 & 1.60 & 3.04 & 3.99\\
\midrule
MMS & 3.48 & 9.84 & 10.78 & 9.90 & 4.85 & 9.29 & 9.35\\
MMS+LOGIT & 3.10 & 9.59 & 9.73 & 8.39 & 4.19 & 8.66 & 8.51\\
MMS+HID & 3.30 & 9.64 & 9.78 & 8.75 & 4.20 & 8.67 & 8.63\\
MMS+COMBO & 2.94 & 9.23 & 9.13 & 8.24 & 3.87 & 8.13 & 8.12\\
\midrule
TA w/o FA fallback & 3.05 & 9.12 & 11.63 & 16.81 & 12.11 & 18.43 & 11.55\\
TA & 3.25 & 9.76 & 10.07 & 9.21 & 4.30 & 8.82 & 8.86\\
TA+LOGIT & \textbf{2.69} & 9.08 & 8.92 & 8.09 & \textbf{3.22} & \textbf{7.41} & 7.88\\
TA+HID & 2.71 & \textbf{9.07} & 8.80 & \textbf{8.03} & 3.55 & 7.78 & 7.86\\
TA+COMBO & 2.71 & 9.11 & \textbf{8.67} & 8.07 & 3.43 & 7.43 & \textbf{7.80}\\
\bottomrule
\end{tabular}
\end{table*}

\section{Experimental Results}
\subsection{Data and experimental setup}

The training data include AISHELL~\cite{bu2017}, WeNetSpeech~\cite{zhang2021}, LibriSpeech~\cite{panayotov2015}, GigaSpeech~\cite{chen2021}, and KeSpeech~\cite{tang2021}. All systems use the same training data. Table~\ref{tab:cer} lists the six evaluation sets.

All systems use a streaming Conformer~\cite{gulati2020}, a temporally downsampling projector, and Qwen2-1.5B~\cite{yang2024}. They are initialized from the same non-streaming ASR-LLM trained on the same Chinese and English speech data. Before training on these data, the non-streaming model's encoder and projector are initialized with pretrained parameters from FireRedASR2S~\cite{xu2026fireredasr2s}. This non-streaming model provides the distillation targets, with its parameters fixed throughout student training. Hidden-state distillation uses LLM layers $\mathcal S=\{7,14,21,28\}$, each with a separate, randomly initialized auxiliary Qwen2 decoder block. The encoder, projector, and LLM are jointly updated without LoRA. We train all models for five epochs on 32 GPUs using Adam with an initial learning rate of $10^{-3}$, a plateau-based learning-rate scheduler, and a gradient clipping threshold of 5. The audio-duration batch budget is set to 90 seconds. The distillation temperature is $\tau=2$. We ablate alignment-path, logit, and hidden-state supervision to assess their contributions.

The audio chunk duration is 240 ms. The text-token limit per chunk is seven in both training and evaluation. All reported variants use this same evaluation configuration. Both FA training labels and evaluation timestamps are generated using the MMS multilingual forced aligner, a CTC-based model trained on 31,000 hours of speech covering 1,130 languages~\cite{pratap2023}. 

In the experiments, MMS denotes systems trained with MMS forced alignments, while TA (teacher-aligned) denotes systems trained with alignment paths derived from the non-streaming teacher, including confidence-based FA fallback. Without a suffix, MMS and TA use neither logit nor hidden-state distillation. TA+LOGIT adds logit distillation, TA+HID adds hidden-state distillation, and TA+COMBO uses both. The same suffixes apply to the MMS variants. The fallback threshold $\theta_{\mathrm{FA}}$ was set to 0.30 based on preliminary experiments on AISHELL-1 and kept fixed for all TA variants across the evaluation datasets. With this threshold, FA fallback is applied to 59.44\% of alignment units across the training data before monotonic search. MMS and TA weights were selected separately (Table~\ref{tab:stream}).

\subsection{Evaluation metrics}
We report character error rate (CER) for Chinese and word error rate (WER) for English. All denotes the aggregate error rate reported across the six test sets.

Flicker measures revisions between consecutive partial hypotheses. Let $H_{u,t}$ be the hypothesis after update $t$ of utterance $u$, with $H_{u,0}=\varnothing$, and $T_u$ its final update. The revision count is
\begin{equation}
r_{u,t}=\left[d(H_{u,t-1},H_{u,t})-\left(|H_{u,t}|-|H_{u,t-1}|\right)_+\right]_+,
\end{equation}
where $d$ is Levenshtein distance, $|H|$ is sequence length, and $(x)_+=\max(0,x)$. We set $r_{u,t}=0$ for updates consisting only of completing the last English word and appending new words. Chinese uses character units and English uses word units. We report micro-averaged flicker:
\begin{equation}
\mathrm{Flicker}=100\%\times\frac{\sum_u\sum_{t=1}^{T_u}r_{u,t}}{\sum_u|H_{u,T_u}|}.
\label{eq:flicker}
\end{equation}

Emission delay is the difference between the output timestamp and the corresponding FA reference endpoint. We report the mean emission delay.

\subsection{Recognition performance}
Without logit or hidden KD, TA improves all six test sets, reducing aggregate error rate from 9.35\% to 8.86\% (5.2\% relative; Table~\ref{tab:cer}). Without fallback, error rises to 11.55\% despite gains on AISHELL-1 and KeSpeech. Degradation on WeNetSpeech and LibriSpeech may stem from complex acoustic conditions and dispersed attention for English function words such as prepositions, respectively, highlighting the need for fallback.

KD improves both MMS and TA. Distilled TA variants achieve 7.80--7.88\%, versus 8.12\% for MMS+COMBO. TA+COMBO is best overall and improves all six sets over MMS+COMBO, with relative reductions of 3.9\% over MMS+COMBO and 16.6\% over undistilled MMS. A substantial gap to the non-streaming teacher remains (7.80\% versus 3.99\%).

\begin{table}[!ht]
\centering
\caption{KD weights and streaming behavior. $\lambda/h$: logit/hidden weights; F: micro flicker (\%); L: mean emission delay (ms).}
\label{tab:stream}
\renewcommand{\arraystretch}{0.92}
\begin{tabular}{lrrrr}
\toprule
System & $\lambda$ & $h$ & F$\downarrow$ & L$\downarrow$\\
\midrule
MMS & 0 & 0 & 3.47 & 129.9\\
MMS+LOGIT & 0.2 & 0 & 2.59 & 124.9\\
MMS+HID & 0 & 0.1 & 3.15 & 130.7\\
MMS+COMBO & 0.2 & 0.5 & \textbf{2.49} & 127.3\\
\midrule
TA & 0 & 0 & 4.01 & 132.2\\
TA+LOGIT & 0.05 & 0 & 3.25 & 126.0\\
TA+HID & 0 & 0.5 & 3.52 & \textbf{124.2}\\
TA+COMBO & 0.05 & 1.0 & 3.18 & 128.3\\
\bottomrule
\end{tabular}
\end{table}

\subsection{Streaming stability and delay}
Within TA, combined KD reduces flicker from 4.01\% to 3.18\% and mean emission delay from 132.2 to 128.3 ms (Table~\ref{tab:stream}). TA+HID achieves the lowest mean emission delay at 124.2 ms, while TA+COMBO has the lowest flicker among the TA variants.

However, all teacher-aligned variants exhibit higher flicker than their forced-alignment counterparts under the corresponding distillation configurations. Without logit or hidden-state KD, TA increases flicker from 3.47\% to 4.01\% relative to MMS. With both distillation losses, TA+COMBO has higher flicker than MMS+COMBO (3.18\% versus 2.49\%), despite similar mean emission delays (128.3 versus 127.3 ms). Thus, distillation improves stability within TA, but the recognition gains over the corresponding MMS systems come with increased flicker.

\section{Conclusions}
We proposed alignment-path distillation to transfer a non-streaming teacher's text--audio alignments to streaming speech recognition. Compared with training using FA, teacher-aligned training achieves relative error rate reductions of 5.2\% without logit or hidden-state distillation and 3.9\% with both losses. The complete framework achieves a 16.6\% relative error rate reduction over FA-based training without either loss. With both losses, teacher-aligned training yields similar mean emission delay but higher flicker than FA-based training.

\end{document}